# Revealing evolutions in dynamical networks

Matteo Morini, Patrick Flandrin, Eric Fleury, Tommaso Venturini, Pablo Jensen[1]
IXXI, ENS de Lyon, INRIA, CNRS, LIP UMR 5668, LP UMR 5672

## 1. Introduction

The description of large temporal graphs requires effective methods giving an appropriate mesoscopic partition. Many approaches exist today to detect "communities", ie groups of nodes that are densely connected (Fortunato, 2010), in static graphs. However, many networks are intrinsically dynamical, and need a dynamic mesoscale description, as interpreting them as static networks would cause loss of important information (Holme and Saramaki, 2012; Holme 2015). For example, dynamic processes such as the emergence of new scientific disciplines, their fusion, split or death need a mesoscopic description of the evolving network of scientific articles.

There are two straightforward approaches to describe an evolving network using methods developed for static networks. The first finds the community structure of the *aggregated* network, ie the network found by aggregating the nodes and their links at all times. However, this approach discards most temporal information, and may lead to inappropriate descriptions, as very different dynamic data can give rise to the identical static graphs (Berger-Wolf and Saia, 2006). To avoid this problem, the opposite approach closely follows the evolutions and builds networks for successive time slices by selecting the relevant nodes and edges. Then, the mesoscopic structure of each of these slices is found *independently* and the structures are connected in various ways to obtain a temporal description (Berger-Wolf and Saia, 2006; G. Palla et al, 2007, Rosvall and Bergstrom, 2010, Chavalarias and Cointet, 2013). By using an optimal structural description at each time slice, this method avoids the inertia of the aggregated approach. Its main drawback lies in the inherent fuzziness of the communities, which leads to "noise" and artificial mesoscopic evolutions, with no counterpart in the real evolutions of the data. For example, rather different partitions have a very close modularity (Good et al, 2010), and minor changes in the network may lead to quite different partitions in successive time slices, which would be inadequately interpreted as major structural changes.

Several methods have been proposed to overcome the problems of these two extreme approaches (Gauvin et al. 2015, Peel and Clauset, 2014, Mucha et al, 2010, Kawadia and Sreenivasan 2012). Here, we present a new approach that distinguishes real trends and noise in the mesoscopic description of social data using the continuity of social evolutions. To be able to follow the dynamics, we compute partitions for each time slice, but to avoid transients generated by noise, we modify the community description at time *t* using the structures found at times *t-1* and *t+1*. We show the relevance of our method on the analysis of a scientific network showing the birth of a new subfield, wavelet analysis. This field represents a difficult test because it has arisen out of the collaboration of several disciplines, producing a rich history, made by many entangled streams.

---
[1] Corresponding author

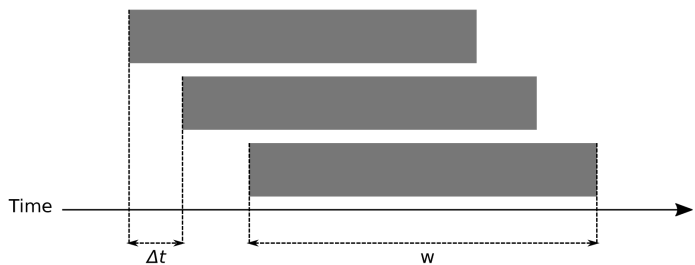

*Figure 1: Sliding temporal windows*

## 2. Method description

Our method consists in four steps:
(1) The dataset is first divided into temporal windows of *w* years, translated by *Δt* years (Figure 1).
(2) In the second step, community detection is carried out *independently* for each window by any method. This leads to a structure that follows as closely as possible the real mesoscale dynamics, at the price of some noise. To selectively delete the noise, while keeping the real evolutions, one has to split or merge communities at each slice, depending on the relations between the successive communities on longer time scales.
(3) For this, the third step systematically computes all the *similarities* between communities at times *t-2, t-1, t, t+1* and *t+2*. For each community at time *t*, we keep *only* the most similar communities at times *t-2* through *t+2*, thus defining its "ancestor" (most similar community at *t-2*) $P_{t-2}$, "predecessor" (at *t-1*) $P_{t-1}$, "successor" (at *t+1*) $S_{t+1}$, and "grandchild" (at *t+2*) $S_{t+2}$. These strong long-term links allow to discriminate real evolutions from noise, by taking advantage of the relative continuity and stability of social evolutions on appropriate time scales. For example, and with regard to the dataset used, a new scientific field does not appear and disappear in a single year.
(4) The fourth (and final) step then uses this long-term information to iteratively select all the time windows and optimize the complexity score (Equation 1). For this, we *merge* communities that appear to be unduly split by the independent community detection (Figure 2a), and *split* communities that appear to be artificial merges (Figure 2c). In practice, we identify artificial merges at time *t* by the links between the "predecessor" communities (at *t-1*) and the "successor" ones (at t+1). If these two are linked (as in Figure 2c), then we assume that these two trends represent the real evolution, and the merge at time *t* arises out of noise in the community detection. We then split the community, attributing the nodes to each of the trends by a simple intersection procedure (for details, see SI, par. 1). In any other case, when there are missing links between the communities (as in Figure 2d), we assume that a real merge has been detected, which is then followed by a split between two different streams. The same procedure is applied to distinguish between real and artificial splits (Figure 2a-b). This procedure goes on as long as there exists an artificial split or merge.

(5) At the end of the procedure, we obtain a description of the network evolution at the mesoscale, the unit of description being now several streams of connected communities. Note, however, that the final description may depend on the set of initial partitions. To render our method robust, we compute a "complexity" score (Equation 1) for different final descriptions and use the one with the highest score, leading to the "richer" story that can be told avoiding noise. The merit of our approach is, by eliminating most of the noise, to limit these complex turbulent regions to the real transformations that should not be discarded: things should be made simple, but not too simple. This score is computed as:

$$C_S = \frac{\sum_{u \in (u_s \cup u_m)} s_u - \sum_{u \in (u_r \cup u_x)} s_u}{\sum_{u \in G} s_u}$$

Where:
  u = nodes in G
  $s_u$ = size of node u
  $u_s$ = resulting split nodes in "structural" (real) splits;
  $u_m$ = resulting merged node in "structural" (real) merges;
  $u_r$ = resulting split nodes in ephemeral (noise) splits;
  $u_x$ = resulting merged node in ephemeral (noise) merges.

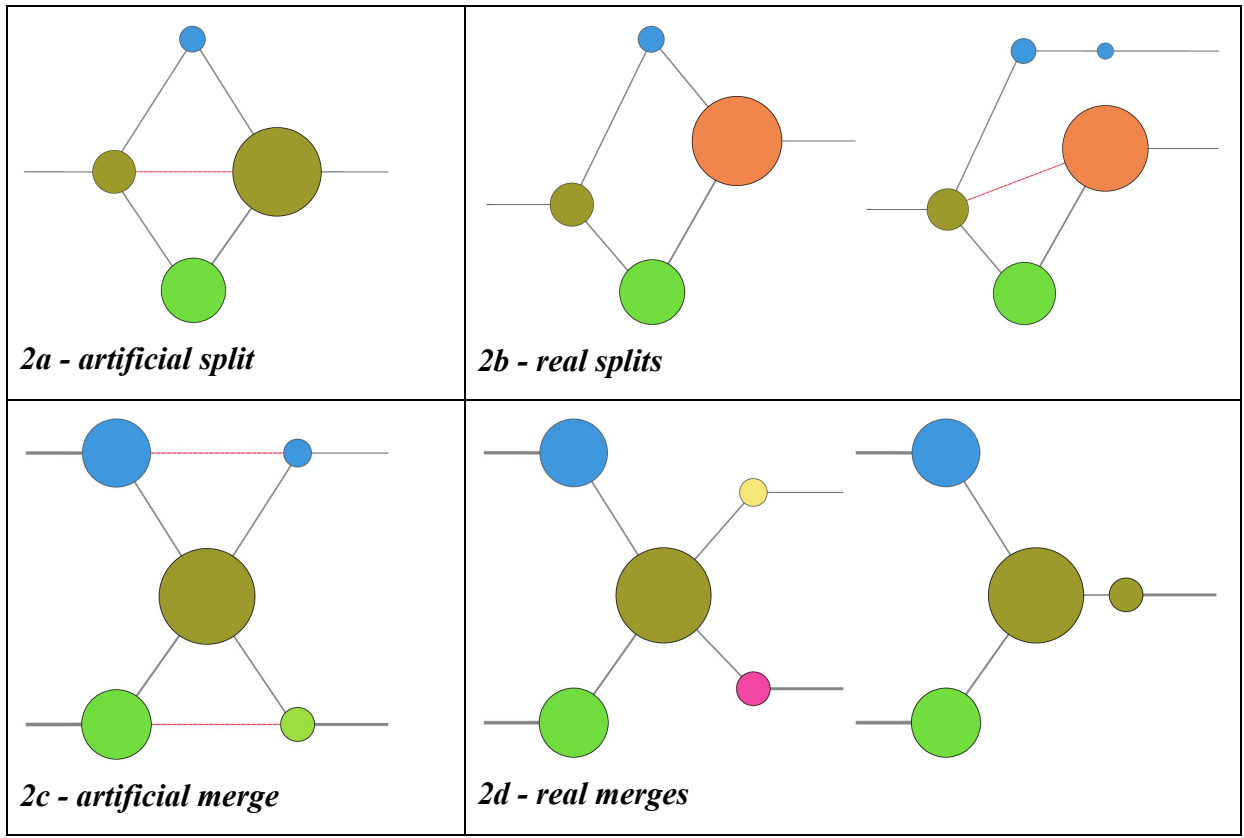

*Figure 2.abcd: Ephemeral (artificial) versus structural (real) events*

To visualize the output, we align the communities that belong to the same "laminar stream", defined as a succession of communities that are all connected, by both t+/-1 *and* t+/-2 links. More formally, a laminar stream LS is defined as an ensemble of communities $C_i$ such as:

$$C_i \in LS, P_{t-2}(C_i) \in LS, P_{t-1}(C_i) \in LS, S_{t+1}(C_i) \in LS, S_{t+2}(C_i) \in LS$$

where $P_{t-1}(C_i)$ is the predecessor of $C_i$ at *t-1*, etc. In some sense, the method tries to produce an output which is as close as possible to a collection of laminar flows, i.e. a set of independent stories. However, real systems are generally more complex, with some "turbulent regions", where real splits produce new streams, flows become intermingled and new subfields are generated. This turns out to be the case in the real case application we describe below.

## 3. Emergence and evolution of a new scientific field: wavelet analysis

We test the method on an evolving network of scientific articles related to the emergence of a new field: wavelets analysis. This technique, developed through collaborations among mathematicians, physicists and electrical engineers, has been fundamental for signal/image processing, leading for example to the well-known jpeg compression format. Wavelets history is interesting as a test case because it is a recently born subfield (seminal paper in 1984), for which robust scientometrics records are available. To define the relevant set of publications, we identified 83 key actors of the early developments of the field. The list was established using expert advice (one of the authors, PF) and bibliographic searches. We then retrieved all their publications (from 1970 to 2012), obtaining 6,500 records from Web of Science. We used 4-years wide time slices ($w=4$), separated by one year ($\Delta t=1$). For each slice, we first defined a network using the articles as nodes, linked by their common references (bibliographic coupling, Kessler 1963, articles sharing less than 2 references are not linked). We then follow the method described above, using maximization of modularity for each slice and the Jaccard similarity index (Jaccard 1901) to compute the similarities between the successive communities. The final result is represented in Figure 3.

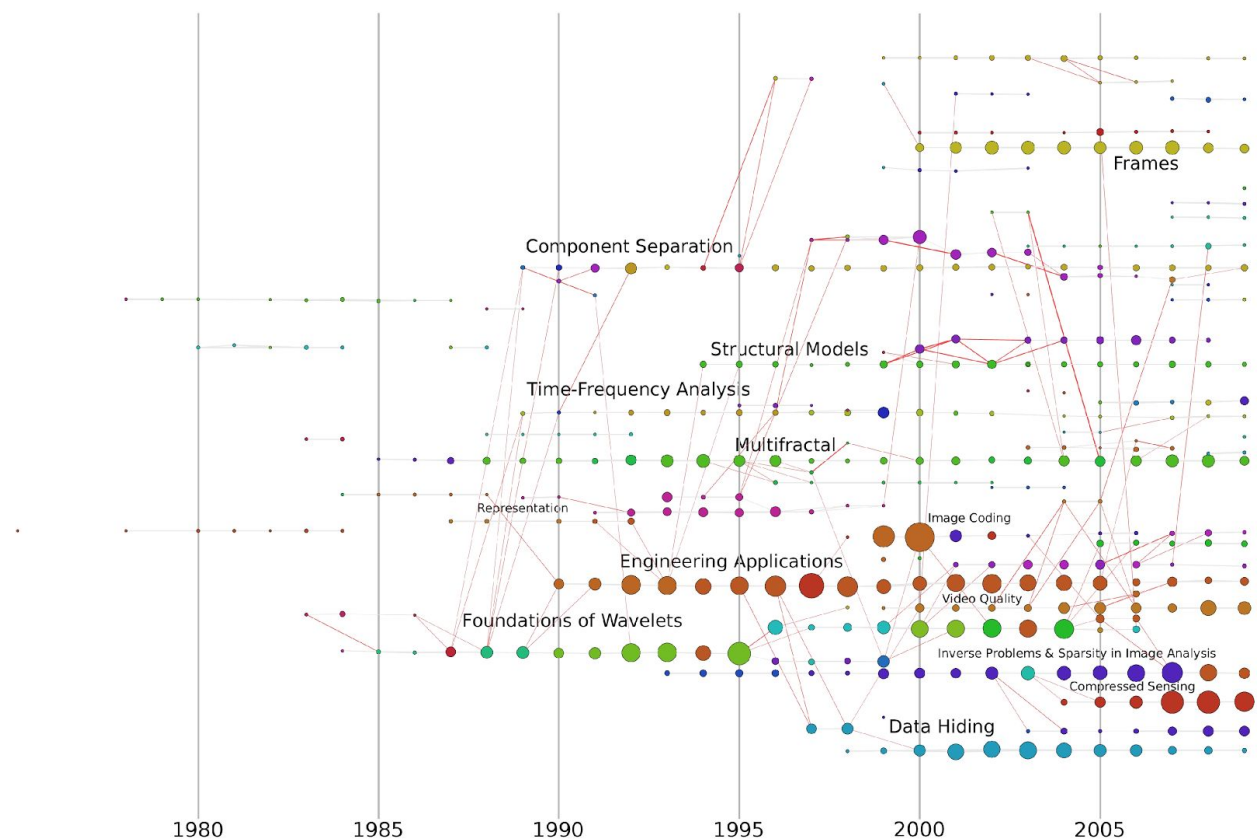

*Figure 3: Overview of the history of wavelets. Streams are labelled according to the subfield within wavelets development (see text for details). Zoomable version available at*
*http://perso.ens-lyon.fr/matteo.morini/wavelets/flows/disciplines.pdf*
*Position of each community: x=year, y=aligned according to streams*

We can now address two important points:
(1) What have we learnt about wavelets evolutions using our method?
(2) Methodological: what do we learn about our method from this example? How important are artificial splits/merges, quantitatively and qualitatively, i.e. to understand the history of wavelets?

## 4. Automatic wavelets history

A first idea about wavelets evolutions can be derived from the evolution of modularity (Figure 5). Roughly speaking, a high modularity value corresponds to isolated clusters, while low values point to highly interconnected networks. The analysis shows that there are three main stages. In an initial phase (before ~1985), researchers work in different, quite unrelated fields and modularity is high (the network of all articles is shown in Figure SI.2). Then, in the 1990s, wavelets appear as a common topic whose use gains momentum, defining a new, specific field that interlinks the publications of our set of authors, leading to a minimum in modularity. After this, modularity increases again, pointing to a new, softer divergence, as the initial levels are not reached. Wavelets become a mature tool, that are less an object of interest *per se*, serving instead a more ancillary role within specialized communities and paving the way for new avenues of research, by developing new tools (as "compressed sensing") or applying wavelets to specific domains, such as Astrophysics images.

Our approach reveals the major structural flows that define the subfields within wavelets development (Figure 3). For each stream, we indicate its name, the main author and the initial/ending dates.
- The stream, "Foundations of wavelets" (1983-2006), starts in 1983 with foundational articles by mathematician Alexis Grossmann. Most wavelets research streams emerge from it, as "Time-frequency analysis" (Flandrin, 1989-2009) and "Component separation" (wavelets without orthonormal bases, Szu, 1989-2009). Starting from mathematical physics, this streams builds wavelets as a rigorous mathematical formalism (80% of its articles are published in Mathematics' journals), but adapted to Engineering concerns.
- In 1990, a central stream, the stem from which most of the subsequent streams will emerge ("Engineering applications", 1990-2009), is created by the fusion of Vetterli's research with a split of the founding stream. This subfield is less concerned by theoretical developments than by practical applications, and most of its articles are published in Engineering journals. The stream "Representation" (Unser, 1986-1992) joins it in 1993, leading to a focus on design. The most important subfields originating in "Applications" are :

- "Inverse problems & sparsity in image analysis" (Starck, 1993-2009), which after focusing on applications on astrophysics images, deals with more general problems in image analysis. It will lead to another important stream, "Compressed sensing" (Baraniuk, 2004-2009)
- "Structural models" (Wilsky, 1994-2009)
- "Image coding", building the theoretical foundations of image coding (Vetterli, 1998-2006)
- "data hiding" (Ramchandran, 1998-2009)
  - Note that there are also some "laminar flows", that interact only peripherally with other lines of research, leading to a linear, simple sequence of communities. Examples of these relatively independent lines of research are the group lead by Alain Arneodo ("Multifractal", 1985-2009), "Frames" (Grochenig, 2000-2009), "Video quality" (Bovik, 1999-2009). These laminar flows represent subfields that apply wavelets to specific objects, without contributing much to the methodological developments.

Finally, it is instructive to look for the position of Yves Meyer, the 2017 prestigious Abel prize for "his pivotal role in the development of the mathematical theory of wavelets". As the number of his publications is not very high, he does not appear explicitly as the main author of any stream. However, his publications are highly cited in the stream "foundations of wavelets", revealing his importance for the mathematical developments. His "pivotal role" of connecting ideas and people, notably in conferences, cannot be seen in our network only made from publications.

## 5. Test of the method

Overall, our method has lead to the split (12) and merge (12) of 24 communities, representing ~10% of all the articles in the database. An example of an important artificial merge detected by our method, similar to the one sketched in Fig 2c, is given in Figure 4. Even if it is clear, looking at the overall history, that there existed two distinct streams of research for 20 years, the independent initial partition merged the communities from these two trends in 1994. Since there exist links (i.e. shared references) among the articles of the two communities (see the articles' network in Fig. SI.1), there is a significant probability that an independent partition algorithm will gather them in a single community. Our method allows to avoid this artificial merge of two distinct streams of research, which belong to different disciplinary traditions, as one subfield is focused on Mathematics, while the other privileges Engineering.

More generally, Figure 5 shows that the rearrangements of partitions demanded by the maintenance of the streams flows leads to negligible losses in the quality of the instantaneous partitions as quantified by modularity. This is important, as it shows that we maintain a close adaptation to the temporal variations, while choosing the partition that best fits the overall evolution.

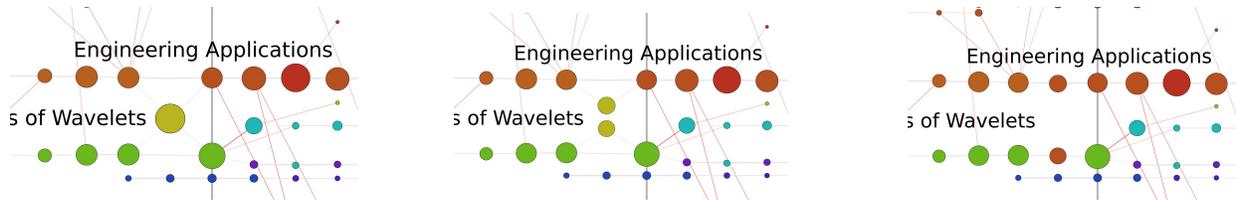

Figure 4: Detail of 1994 split. From left to right, the unduly merged communities are split and assigned to their respective streams.

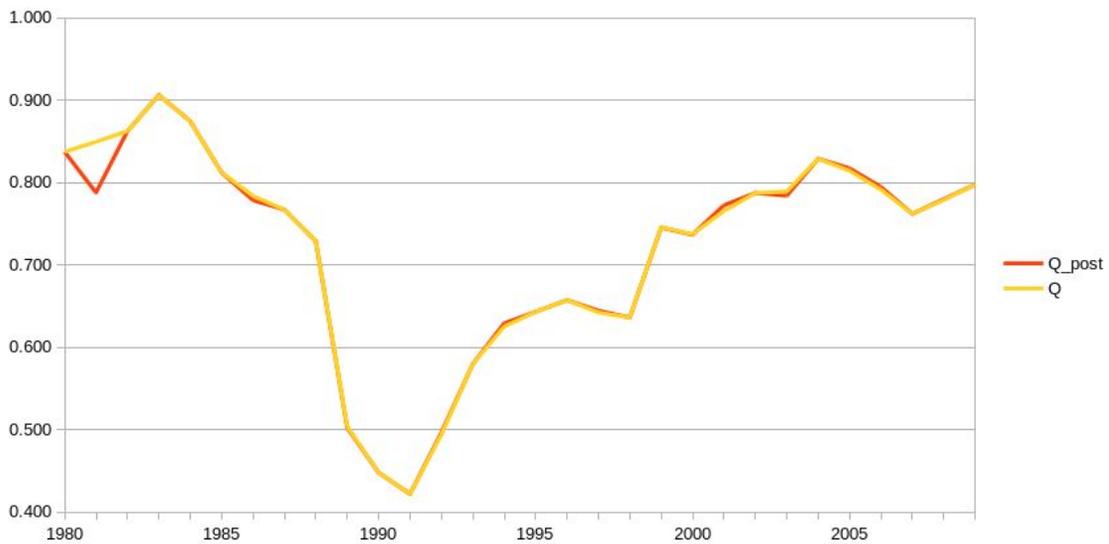

Figure 5: Modularity evolution, initial and final.

There is a clear minimum in modularity for year 1991, pointing to a homogeneous network without much structure, which results from the effective mixing of different disciplinary traditions around the new object (wavelets).

## 6. Comparison to other methods

Our approach offers decisive advantages over existing methods:

- It handles naturally networks in which nodes appear or disappear at each time step, which is impossible or cumbersome for other methods.
- Contrary to generative models (Jacobs and Clauset 2004; Peixoto and Rosvall 2016, Xu, Kliger and Hero 2014), we do not need to define an *a priori* network community structure (for example, a block model) which may be unadapted to the data.
- Claveau and Gingras (2016) have studied the history of economics using scientometrics data and an approach quite similar to ours. However, to determine the partition at time *t*, they initialize the Louvain algorithm with the partition obtained for the preceding time step. Their approach is therefore limited to partitioning by this algorithm. Moreover, the authors do not justify why their approach represents a sound way of adding some inertia to real-time partitioning.

## 7. Concluding remarks and future work

To make sense of transformations, we need evolving categories that can, at the same time, readily adapt to the changes and maintain the continuity of the description. Our method starts from the idea that the unity of an evolving social process rests on the continuity of its transformations, and uses the available mid-term temporal information to reveal structural trends from noisy data, without the assumption of an *a priori* community structure. It can be adapted to any partitioning method and to any similarity measure between communities at different times. Used on scientific data, our method automatically produces a rich historical account, an objective raw material to be discussed by science historians.

There is much room for improvement. The relevant time scales (*w*, delta t) have to be chosen from expert knowledge, and we cannot deal with real-time data, as we use the future to infer the best present partition. We now work to introduce, through a hidden Markov model, an explicit meso temporal scale at which transformations (splits/merges) are supposed to happen for a pair of streams.

**Addendum: List of WLTC selected authors**

Aldroubi, A; Antoine, JP; Antonini, M; Argoul, F; Arneodo, A; Auscher, P; Bacry, E; Baraniuk, RG; Barlaud, M; Battle, GA; Benveniste, A; Beylkin, G; Bijaoui, A; Boudreaux-Bartels, GF; Bovik, AC; Burrus, CS; Chui, CK; Cohen, A; Coifman, RR; Compo, GP; Dahmen, W; Daubechies, I; Devore, RA; Donoho, DL; Elezgaray, J; Escudie, B; Farge, M; Feauveau, JC; Feichtinger, HG; Flandrin, P; Frisch, U; Froment, J; Gopinath, RA; Grochenig, K; Grossmann, A; Guillemain, P; Haar, A; Healy, DM; Heil, C; Herley, C; Holschneider, M; Jaffard, S; Jawerth, B; Johnstone, IM; Jones, DL; Kadambe, S; Kaiser, JF; Kerkyacharian, G; Kronland-Martinet, R; Lawton, W; Lemarie-Rieusset, PG; Lu, J; Lucier, BJ; Mallat, S; Mendlovic, D; Meyer, Y; Micchelli, CA; Morlet, J; Murenzi, R; Muschietti, MA; Muzy, JF; Paul, T; Perrier, V; Picard, D; Ramchandran, K; Resnikoff, HL; Rioul, O; Saracco, G; Shao, XG; Shapiro, JM; Slezak, E; Starck, JL; Szu, HH; Tchamitchian, P; Torresani, B; Unser, M; Vaienti, S; Vetterli, M; Walter, GG; Weaver, JB; Wickerhauser, MV; Willsky, AS; Wornell, GW.

**Supplementary information**

1. Re-splitting unduly merged communities

Occasionally, and because of the inherently noisy community detection process, groups of nodes (articles) can be ambiguously assigned to either two distinct communities, or a single, larger community. When we weigh in the additional temporal information from *t-1* and *t+1*, and observe that the ambiguity can be resolved (e.g. two consistently distinct streams being joined for one step only), we assume that the unduly merged communities can be split back. In order to preserve continuity, each article belonging to the wrongly merged community $C_0$ is assigned to one of two new communities, $C_{01}$, $C_{02}$. We define two sets of articles, $U_a$ and $U_b$, one for each of the two streams, $S_a$ and $S_b$, which correspond to the union of nodes appearing within each couple of predecessor/successor: $P_{a,t-1}$ and $P_{a,t+1}$, and $P_{b,t-1}$ and $P_{b,t+1}$ respectively. Nodes from $C_0$ which belong to the set $U_a$ are assigned to $C_{01}$; similarly, nodes belonging to $U_b$ end up into $C_{02}$. The intersection of $U_a$ and $U_b$ is not necessarily an empty set: because of the fuzziness of communities, a node can appear both in $P_{a,t-1}$ and $P_{b,t+1}$; in this case, and for lack of a ground truth, assignment is performed randomly.

2. Network structure examples, colors according to community

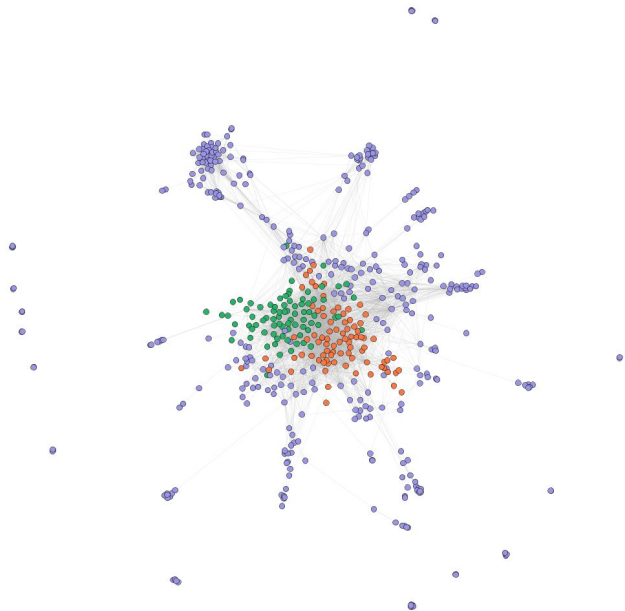

*Figure SI.1, year 1994: artificial (ephemeral) merge happens on a group of nodes (see main text)*

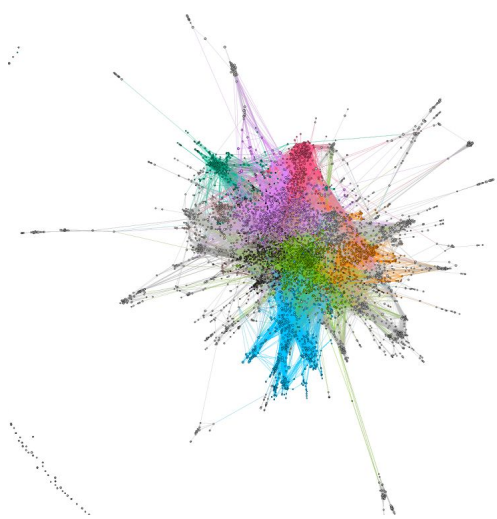

*Figure SI.2, full articles network*

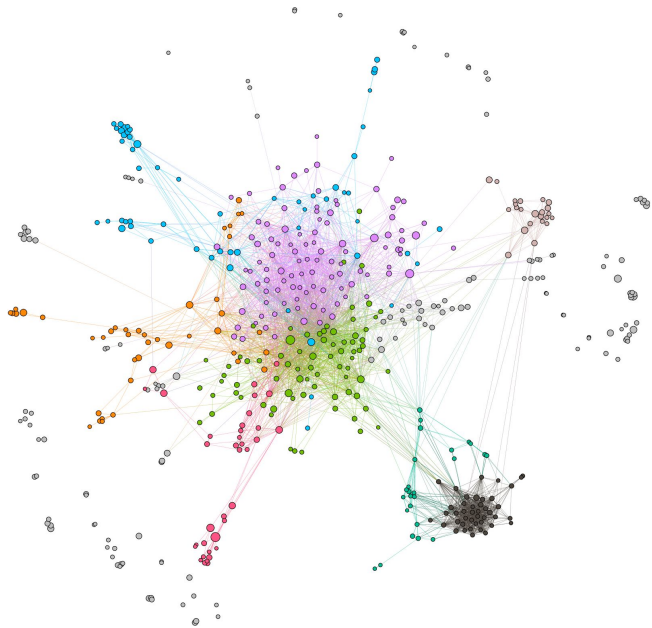

*Figure SI.3, articles network, time window 1994-1997*

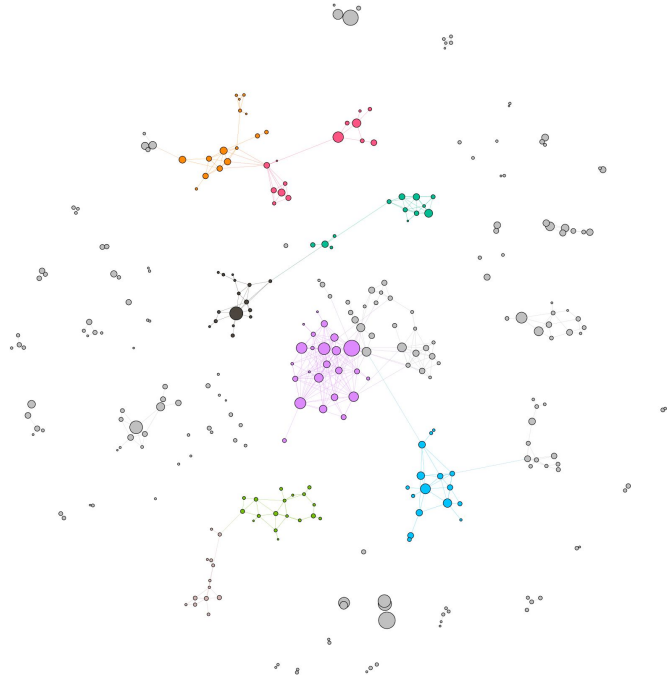

*Figure SI.4, full articles network, up to 1982*